\documentclass[preprint,3p,12pt]{elsarticle}

\usepackage{amssymb}
\usepackage{amsmath}

\journal{}

\begin{document}

\begin{frontmatter}



\title{An Origami-Inspired Energy Absorber}


\author[aff1,aff2]{Shadi Khazaaleh}
\author[aff3,aff4]{Ahmed S. Dalaq}
\author[aff1,aff2]{Mohammed F. Daqaq\corref{cor1}}

\affiliation[aff1]{organization={Department of Mechanical and Aerospace Engineering, Tandon School of Engineering, New York University},
            city={Brooklyn},
            postcode={11201},
            state={NY},
            country={USA}}
            
\affiliation[aff2]{organization={Engineering Division, New York University Abu Dhabi (NYUAD)},
            addressline={Saadiyat Island}, 
            city={Abu Dhabi},
            country={United Arab Emirates}}
            
\affiliation[aff3]{organization={Bioengineering Department, King Fahd University of Petroleum $\&$ Minerals},
            addressline={31261}, 
            city={Dhahran},
            country={Saudi Arabia}}

\affiliation[aff4]{organization={Interdisciplinary Research Center for Health and Biosystems, King Fahd University of Petroleum $\&$ Minerals},
            addressline={31261}, 
            city={Dhahran},
            country={Saudi Arabia}}

\cortext[cor1]{Corresponding Author: mfd6@nyu.edu}

\begin{abstract}
The design of effective and compact energy absorption systems is key to the survivability and durability of many man-made structures and machines. To this end, this work presents the design, assessment, and implementation of a novel origami-inspired energy absorber that is based on the Kresling origami pattern. The absorber consists of a Kresling origami column positioned between the loading point and an energy dissipation module. By exploiting its unique inherent translation-to-rotation coupling feature, the primary function of the Kresling column is to transmit uniaxial incident loads (shock or impact) into localized rotational energy that can then be dissipated in a viscous fluid chamber. The proposed system has several unique advantages over traditional designs including the ability to $i)$ dissipate energy associated with both torsional and uniaxial loads, $ii)$ control the rotational velocity profile to maximize energy dissipation, and $iii)$ customize the restoring-force behavior of the Kresling column to different applications. Furthermore, the proposed design is more compact since it can realize the same stroke distance of the traditional translational design while being considerably shorter. Through extensive computational modeling, parametric studies, and experimental testing, it is demonstrated that the proposed design can be optimized to absorb all the imparted energy; and out of the absorbed energy, around 40\% is dissipated in the viscous fluid, while the rest is either dissipated by the viscoelasticity of the origami column or stored in it as potential energy. 

\end{abstract}




\end{frontmatter}


\renewcommand{\thefootnote}{\fnsymbol{footnote}}

\section{Introduction}
\label{sec:intro}

The design of effective and compact energy absorption systems is key to the survival of many biological species and to the durability of man-made systems. To achieve this objective, nature has evolved different shock/energy absorbing architectures that typically combine cellular structures with viscoelasticity (e.g., polysaccharides)  \cite{barthelat2016structure, san2020review, zhang2022dynamic}. The use of buffer structures and/or materials to protect the skeleton and internal organs of animals is widespread, ranging from tendons that store elastic energy during shocks \cite{roberts2013tendons}, to the viscoelastic fatty layers found in nearly all animals \cite{grear2018mechanical, arumugam1994effect}. 

\begin{figure}[t!]
\centering
\includegraphics[width=1 \linewidth]{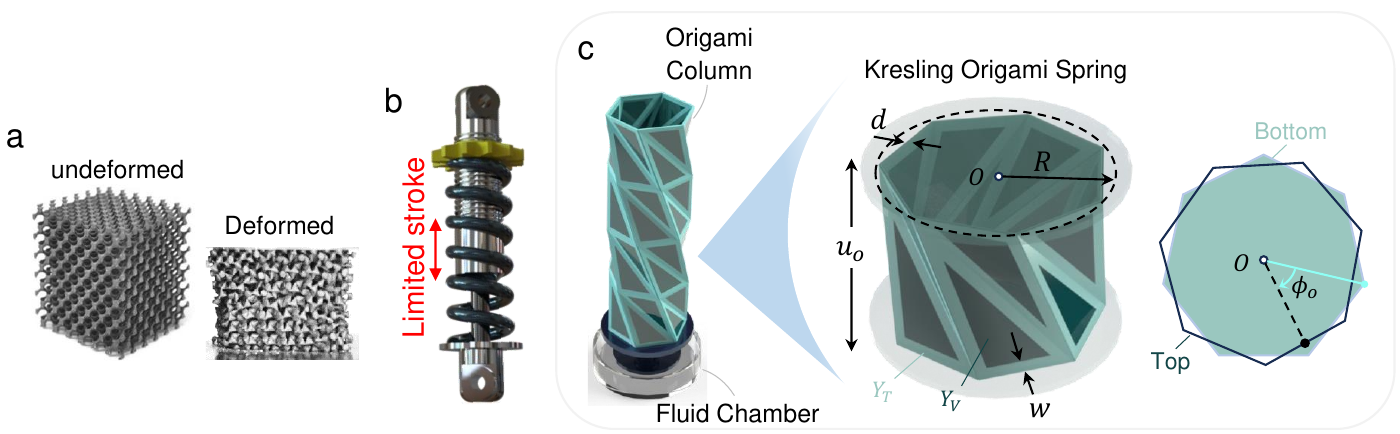}
\caption{(a) Example of energy absorption/dissipation through the crushing of cellular materials. (b) Shock absorber in a vehicle. (c) An origami based device that dissipate energy via a fluid chamber. The system is made from an array of Kresling origami springs with a compliant frame and a rigid core embodied by Tango and Vero 3D printing materials having, respectively, elastic moduli of $Y_T$ and $Y_V$.}
\label{fig:intro}
\end{figure}

Depending on the application at hand, engineered systems may incorporate single- or multi-use shock/energy absorption mechanisms. Single-use mechanisms are typically cheap and rely on damage, buckling, plastic deformations, and/or internal friction of cellular structures to absorb and dissipate energy \cite{shaw1966plastic, abramowicz1984dynamic}. Common packaging materials such as foam and corrugation-core sandwich structures as well as cellular architectured materials that absorb and simultaneously dissipate input energy can be placed in this category (Fig.~\ref{fig:intro}a) \cite{abou2019mechanical, abou2020mechanical, habib2018fabrication, zhang_3d_2021}. Such cellular forms come in various architectures arranged in different configurations, ranging from random Kelvin foams \cite{warren1997linear} to the triply periodic minimal surfaces (TPMS) \cite{al2019multifunctional, tran2021triply, dalaq2016mechanical, dalaq2016finite}.

In applications where a structure or a machine is to be protected against constant shocks and impacts, multi-use energy absorbers must be incorporated. The most famous design is that of the hydraulic shock/energy absorber used in most vehicular systems and machines. As shown in Fig. \ref{fig:intro}b, the system consists of a piston-cylinder arrangement surrounded by a traditional helical spring. The cylinder contains a viscous fluid to dissipate a portion of the imparted energy, while the spring serves to restore the original position of the protected structure  by storing and releasing potential energy. 

In this paper, we propose an alternative approach to the design of the traditional hydraulic energy absorber, which we believe can offer several advantages over the current design. Before we state such advantages, we describe the proposed concept. In the new design shown in Fig. \ref{fig:intro}c, the traditional helical spring is replaced by an origami-inspired column based on the Kresling origami pattern \cite{Kresling2008, Jianguo2015}, while the piston cylinder arrangement is replaced by a chamber that contains a viscous fluid and a disc, which in turn, is connected to the base of the origami column and is free to rotate in the viscous fluid. 

\begin{figure}[t!]
\centering
\includegraphics[width=0.3 \linewidth]{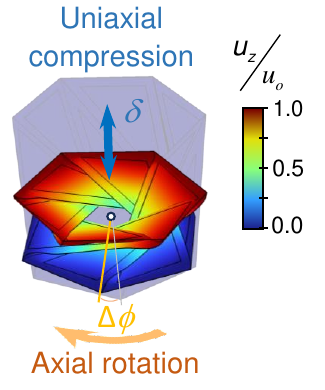}
\caption{Kinematic of a single Kresling spring.}
\label{fig:kinematicsingle}
\end{figure}

To explain the working principle of the new energy absorber, one has to first understand the unique kinematics of the Kresling origami spring and how it provides the restoring force. As shown in Fig. \ref{fig:intro}c, the spring is a cylindrical bellow-type structure consisting of similar triangular panels arranged in cyclic symmetry. One side of these panels forms two $n-$sided parallel polygonal end planes each subscribing a circle of radius, $R$, at the top and bottom faces of the bellow. When a bellow of initial height, $u_0$, and an initial angle $\phi_0$ (between the upper and lower polygons) is subjected to a uniaxial compressive load at one end, the geometric constraints are such that the applied force gets transformed into a torque on the other end, which causes the disc to rotate by an angle $\Delta\phi = \phi - \phi_o$. As such, the translational energy on one end is transformed into rotational energy on the other (Fig. \ref{fig:kinematicsingle}), which can then be dissipated as heat through the shear (frictional) force between the viscous fluid in the chamber and the rotating disc. Once the force is removed, the origami bellow, which is typically constructed using viscoelastic materials, springs back to its initial configuration therewith providing the restoring force behavior that is typically provided by the helical spring in the traditional shock absorber.

This approach offers several advantages. First, this axial-rotational origami mechanism can also transform torsional motions on one end to translational motions on the other,  which, in principle, also permits dissipating torsional loads when the traditional purely axial mechanism cannot. Second, for the same input velocity profile of the impactor, the rotational velocity profile of the disc can be controlled by changing the geometry of the origami column, which permits full control over the amount of dissipated energy. Such velocity control cannot be easily realized when using the traditional shock absorber. Third, the qualitative behavior of the Kresling spring can be finely tuned through a few geometric parameters to encompass linear, nonlinear, quasi-zero-stiffness, and bi-stable restoring force behavior \cite{RmasanaKIOS19, MasanaKIMS, DALAQ2022110541, dalaq_origami-inspired_2023, KHAZAALEH2022109811}. This may offer solutions for applications requiring specific restoring profiles. Finally, the new system is more compact than the traditional translational design because, upon impact, the translational motion of the origami spring is transformed into rotational motion. Thus, the new design can realize similar stroke distance while being considerably shorter.

In this paper, we design, optimize, and assess the performance of this energy absorber both computationally and experimentally. To this end, the rest of the paper is organized as follows: In Sec. \ref{sec:concept}, we describe the origami spring, and its fabrication, then develop a computational model to study its kinematics under impact loading. Subsequently, we develop an experimental testbed and use it to validate the computational model. In Sec. \ref{sec:energy}, we describe the process of calculating and comparing the energy dissipation efficiency across different designs. In Sec. \ref{sec:parametric}, we present a parametric study of the effect of the impact loading and damping on the dissipation efficiency. In Sec. \ref{sec:chamber}, we present and describe the experimental energy dissipation module that was built to assess the efficacy of the proposed system. In Sec. \ref{sec:results}, we build the proposed origami-inspired energy absorbers and present the experimental results for performance evaluation. In Sec. \ref{sec:conc}, we present the important conclusions.

\section{Modeling and Experimental Validation}
\label{sec:concept}
Due to the modular nature of Kresling springs, they conveniently serve as buffer modules that can be positioned between the impact zone and the energy dissipation process. Their primary function is to transmit the incident kinematics and kinetics to an energy dissipation module. The unique inherent translation-rotation coupling ensures a seamless conversion of uniaxial translation into localized, compact rotations in response to any random or volatile incident loads. Given recent understanding of the mechanics and dynamics of Kresling springs and their behavior(s) \cite{DALAQ2022110541,KHAZAALEH2022109811}, we can effectively employ them and precisely control their response in terms of force-deflection and deflection-time by manipulating the parameters introduced earlier in Fig~\ref{fig:intro}c.

One of the key elements for a successful design is the ability to manufacture a reliable spring that is capable of withstanding larger incident loads and be resistant to fatigue. Despite the initial challenges encountered in creating functional and reliable prototypes of the Kresling springs, various research works led to the development of metallic and 3D-printed versions which are suitable for engineering applications \cite{dalaq_origami-inspired_2023, KHAZAALEH2022109811, zhai2018origami, melancon2022inflatable}. Here, we adopt the  approach we proposed in Ref. \cite{KHAZAALEH2022109811}, where the Kresling spring is constructed using a two-phase material composite; a soft frame and rigid core panels that maintain the structural stiffness and geometric characteristics of the Kresling's design. In addition to the geometric parameters, this composite structure introduces two elastic moduli, representing the soft and stiff phases; which are Tango, with a Young's modulus $Y_T$, and Vero, with a Youngs modulus, $Y_V$, respectively (see \ref{app:fab} for details on the fabrication process).

\begin{figure}[t!]
\centering
\includegraphics[width=1 \linewidth]{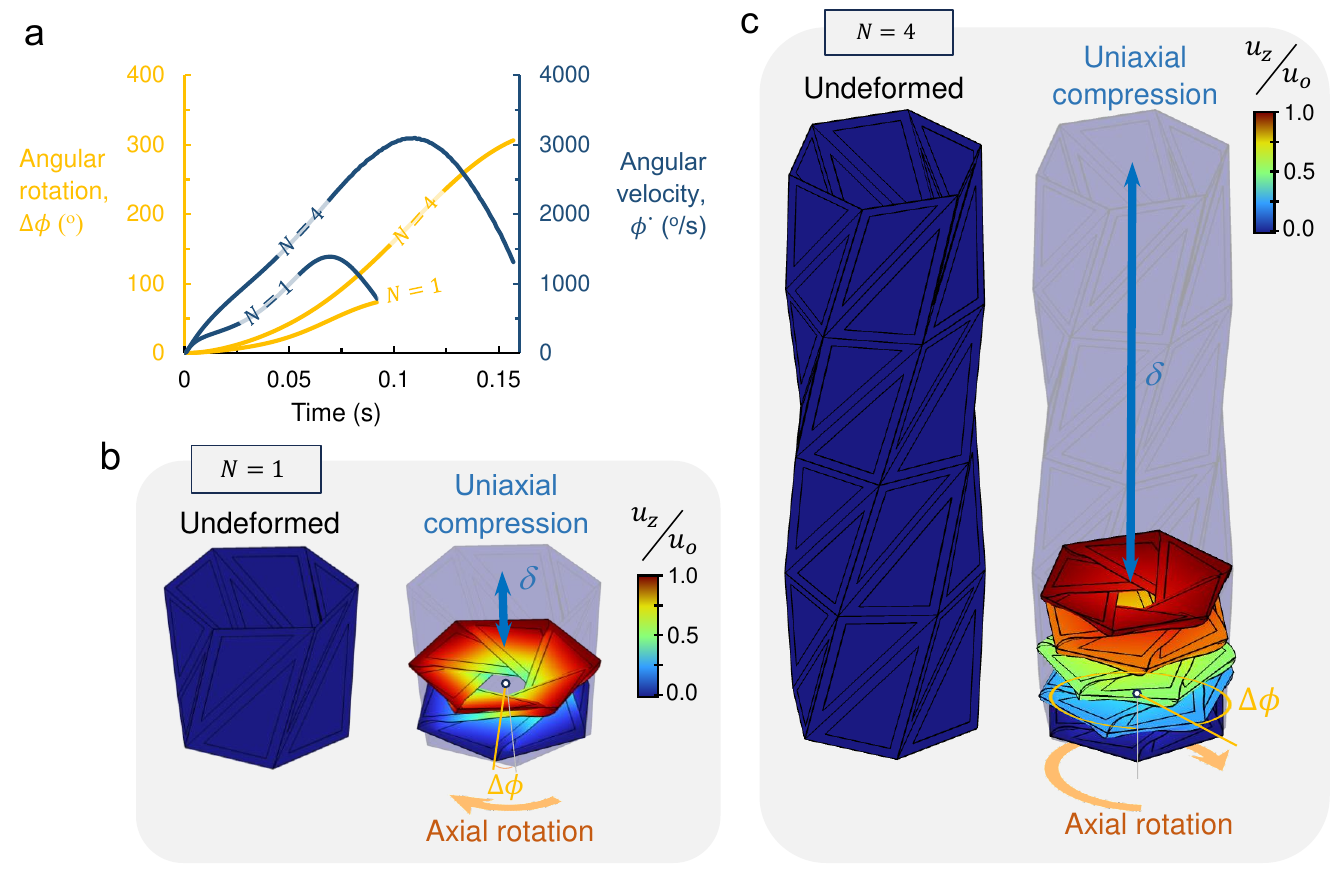}
\caption{The origami column. (a) Time evolution of the angular rotation and velocity for $N=1$ and $4$. (b) Uniaxial compression of a single unit, and (c) 4 units column.}
\label{fig:mechanism}
\end{figure}

To model the kinematics of the spring under loading, we developed and employed the quasi-static computational model based on the work in Ref. \cite{DALAQ2022110541}, but generalized it to take into account dynamic effects (see \ref{app:comp} for details on the computational model).  The computational model accepts generic geometric inputs: $(n, R, u_o, \phi_o)$, and the number of Kresling spring units, $N$, forming the column shown in Fig. \ref{fig:mechanism}c. A known mass, $M$, of 0.95 kg released on top of the Kresling spring drives the compression of the spring under dynamic settings. The selected mass level is sufficient to induce full compression. Self-contact between panels is disregarded in the simulation. The computational model calculates the kinematics, including angular rotation, $\Delta\phi$, and angular velocity, $\dot{\phi}$, over the simulation time. The number of sides is fixed to $n=6$ throughout this study. 

Figure \ref{fig:mechanism}a shows time variation of both $\Delta\phi$ and $\dot{\phi}$ for two examples: $N=1$ and $N=4$. For a single unit ($N=1$), Fig. \ref{fig:mechanism}a and \ref{fig:mechanism}b show how the applied compression, $\delta$, results in a rotation, $\Delta\phi$, at the base due to the kinematic constraints of the Kresling origami. This rotation profile is completely determined by the design parameters $(u_o/R, \phi_o)$. A single Kresling spring unit with parameters: $(u_o/R=1.6, \phi_o=55^{\circ})$ can rotate up to $73^{\circ}$ with a peak angular velocity of $1391^{\circ}/s$ at around 0.07 s. The peak angular velocity appears to consistently occur at $\sim 0.35Nu_o$; that is at simulation time of 0.07s and 0.11s for $N=1$ and $N=4$, respectively. After which there is a rapid decrease in $\dot{\phi}$, towards full rest. Increasing the number of units from $N=1$ to 4 induces $\times4.2$ the angular rotation (from $\Delta\phi=73.6^{\circ}$ to $\Delta\phi=306^{\circ}$), and doubles ($\times2.22$) the angular speed. 

\begin{figure}[t!]
\centering
\includegraphics[width=1 \linewidth]{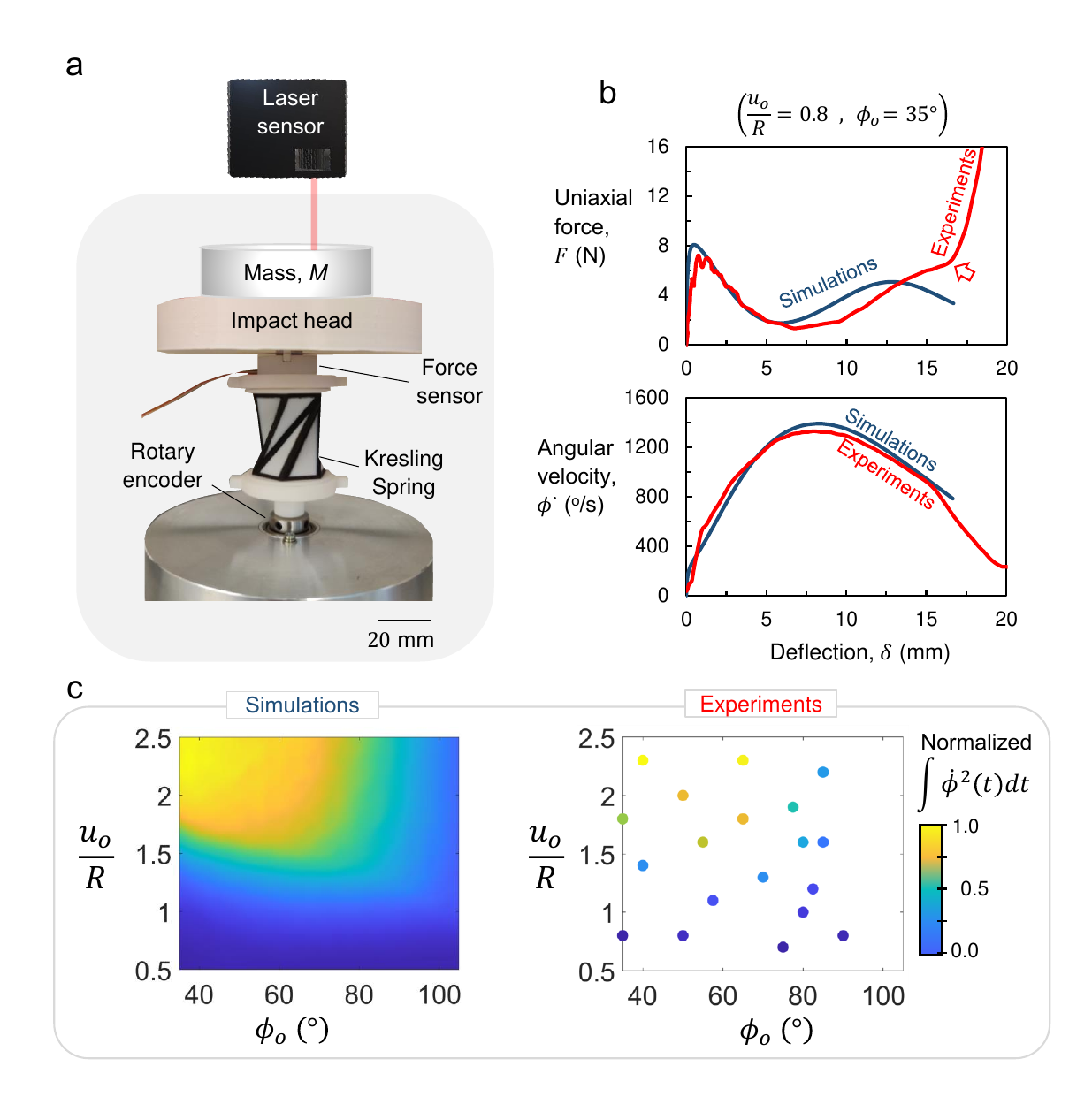}
\caption{(a) Experimental apparatus. (b) Comparison of simulations (in dark blue) with experimental (in red). The arrow points to the instance of densification. (c) Side by side comparison between simulation results and discrete points from experiments across the parameter design space: $u_o/R=[0.5,2.5]$ and $\phi_o=[35^{o}, 105^{o}]$.}
\label{fig:validation}
\end{figure}

To validate the theoretical findings of the dynamic computational model, we developed the experimental setup shown on Fig. \ref{fig:validation}a. The setup is composed of a rotational encoder on which the Kresling spring is fitted with a rigid adapter, measuring the real-time rotational angle $\Delta\phi$ during testing. The top surface of the Kresling spring is fitted with an impact head loaded with a mass, $M$. A sensitive, low-profile force sensor is housed within the rigid adapter that connects the Kresling spring with the impactor, measuring the reaction force, $F$, during compression. A laser sensor is positioned above the impact head and is used to track the real-time position of the mass during its release. The test starts by releasing the mass, then recording the corresponding force-deflection ($F-\delta$) of the spring and the rotational velocity of the rotating base $\dot{\phi}$, as shown in Fig. \ref{fig:validation}b (see \ref{app:exp} for details on the experimental setup). Figure \ref{fig:validation}b also reports the corresponding simulation of the previously described test, using identical mass and material properties. Results demonstrate very good agreement up to the point of experimental densification of the material, which is not taken into account in the computational model.

\section{Energy Dissipation Efficiency}
\label{sec:energy}
When a mass is dropped on top of the Kresling spring, the imparted energy will be absorbed by the structure in different forms, in accordance to the following energy conservation relation:
\begin{equation}
E_{in} = U + U_v + E_v + E_t + E_{losses}~,
\label{eq:energies}
\end{equation}
where $E_{in}$ is the input energy to the system, which can be described as the potential energy of the dropped mass. Given that the dropped height is $Nu_o$, $E_{in}=MgNu_o$, where $g$ is the gravitational acceleration. The rest of the parameters in the equation are such that $U$ is the elastic strain energy stored in the constituent materials of the Kresling spring, $U_v$ is the dissipated viscoelastic part of the strain energy, $E_v$ is the dissipated energy by the interaction between the rotating disc and the viscous fluid (note that the disc has negligible rotary inertia), $E_{losses}$ represents the energy lost to the surroundings, such as by air drag, friction, elastoacoustics, etc, and $E_t$ is the rebound energy after impact.  

When employing a viscous damper, the rotational energy dissipation, $E_v$, is proportional to $\dot{\phi}$ through the relationship $E_v=\int{C_\theta\dot{\phi}^2(t) dt}$, where $C_\theta$ is the rotational damping coefficient. The goal is to maximize this quantity as it is the energy to be dissipated by the viscous fluid. For the purpose of comparing different designs, we assign $C_\theta$ a value of unity ($C_\theta = 1$ Nms/rad). In Fig. \ref{fig:validation}c, we report a contour plot of simulated values of $E_v$ across the parameter design space: $(u_o/R,\phi_o)$ between $u_o/R=[0.5,2.5]$ and $\phi_o=[35^{\circ}, 105^{\circ}]$. To the right, we show the corresponding experimental results for comparison, revealing nearly identical distributions and values. This affirmation allows us to confidently proceed with in-depth design analysis using the developed computational model.

\begin{figure}[t!]
\centering
\includegraphics[width=1 \linewidth]{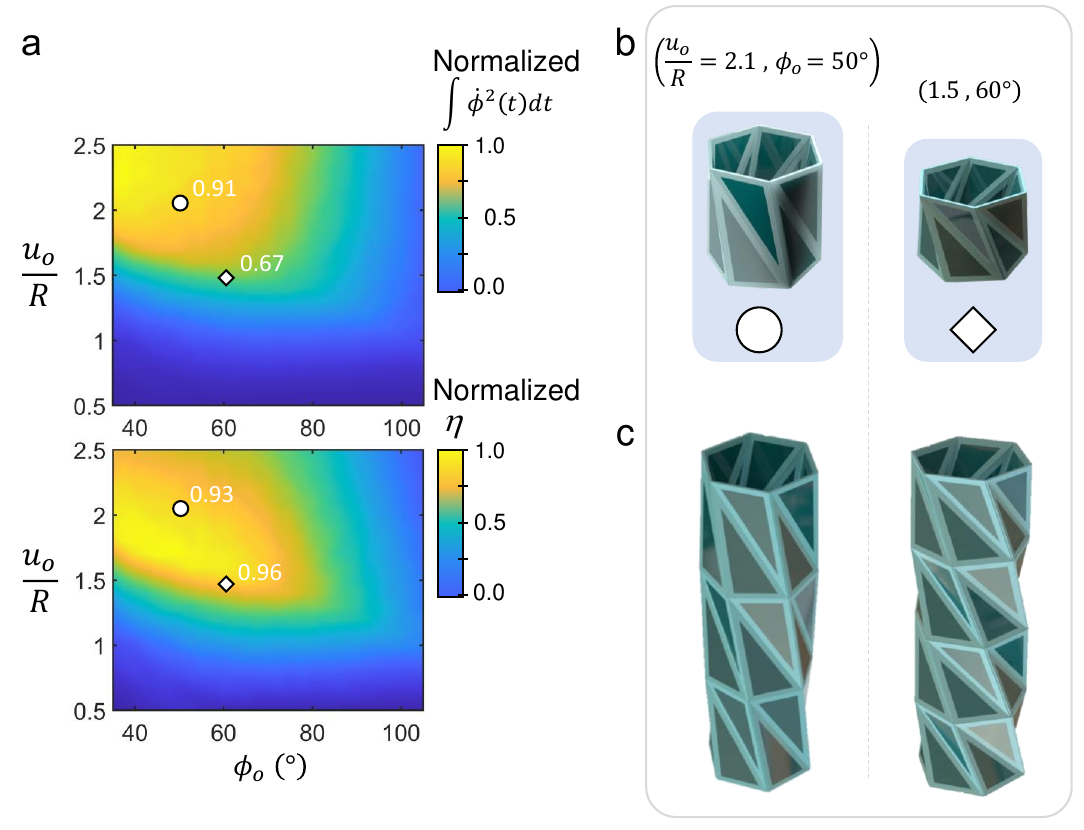}
\caption{(a) Normalized $\int{\dot{\phi}^2 dt}$ and efficiency $\eta$ across $u_o/R=[0.5,2.5]$ and $\phi_o=[35^{o}, 105^{o}]$. Two specific designs are marked with circular and diamond markers as examples of, respectively, a highly energy-dissipative design versus efficient ones. (b) Illustration of Kresling springs of these two design cases: $(2.1, 50^\circ)$ and $(1.5, 60^\circ)$. (c) Demonstration of a stacked array of these designs forming columns of equivalent height but varying efficiency.}
\label{fig:efficiency}
\end{figure}

The efficiency of the dissipation process is simply the ratio between $E_v$ and $E_{in}$; that is 
\begin{equation}
\eta=\frac{E_v}{E_{in}}.
\label{eq:efficiency}
\end{equation}
Since for a single spring unit $E_{in}=Mgu_o$, the efficiency, $\eta$, can be maximized by achieving highest angular velocities over a wide range of $\phi$ while maintaining the lowest vertical height of the stack represented by $u_o$. Taller units, with relatively large $u_o$, clearly have more room for providing rotation at the base than shorter ones. For instance, while the integral values, as described earlier, for the taller design $(u_o/R=2.1, \phi_o=50^{\circ}$, denoted by a circular marker on Fig. \ref{fig:efficiency}a and \ref{fig:efficiency}b) is higher than the shorter one $(1.5, 60^{\circ}$, denoted by a diamond marker on Fig. \ref{fig:efficiency}a and \ref{fig:efficiency}b), the shorter design is relatively more efficient than the taller one. Therefore, when stacked in a column of the same height, as shown in Fig. \ref{fig:efficiency}c, the column made from the shorter, more efficient design yields greater energy dissipation, as will be demonstrated experimentally in the upcoming experimental section.

\section{Parametric Study}
\label{sec:parametric}
\begin{figure}[t!]
\centering
\includegraphics[width=0.9\linewidth]{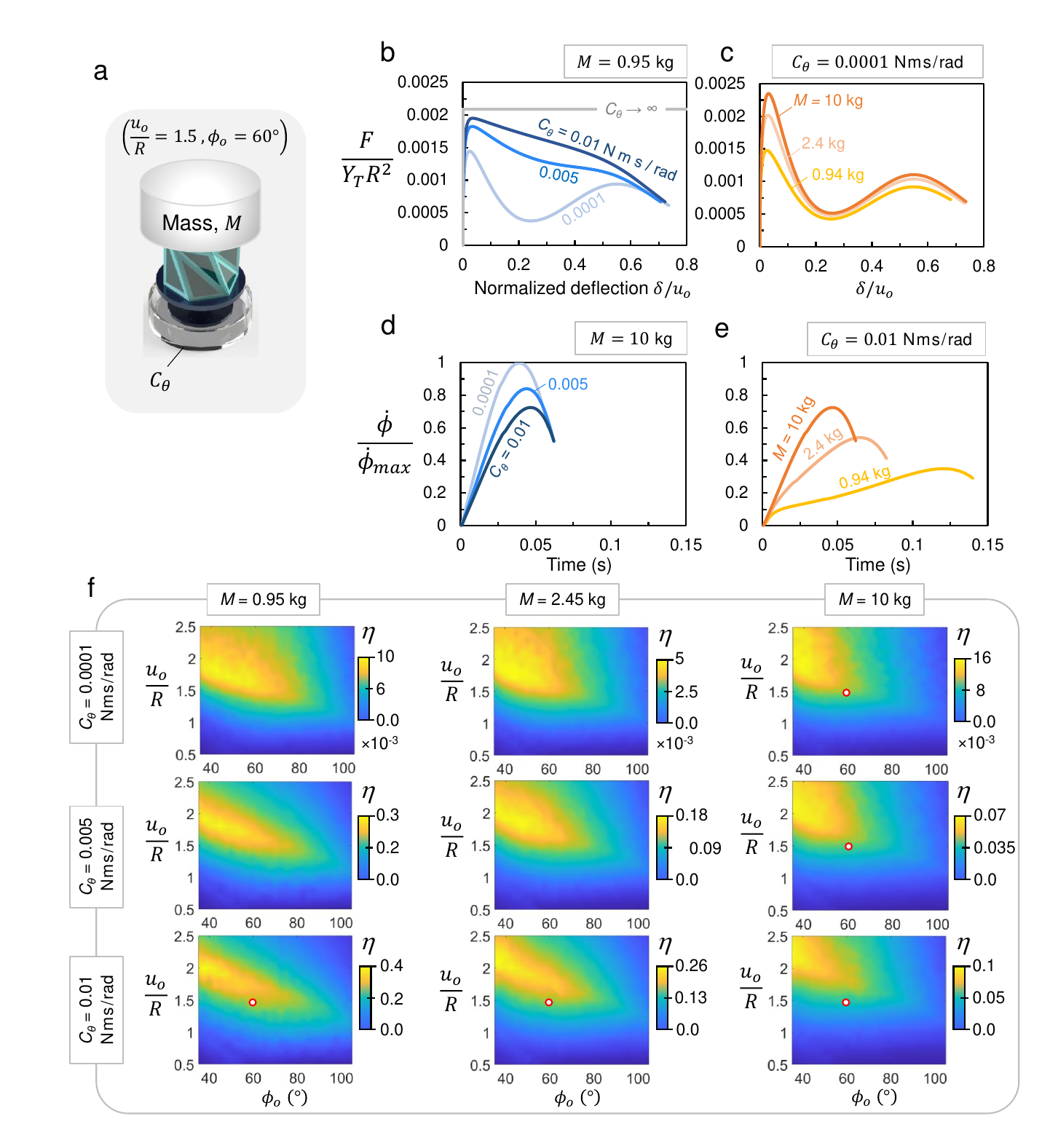}
\caption{(a) Model of a Kresling spring unit with an attached rotational viscous damper used in computational simulations. Simulation results of normalized force, $\frac{F}{Y_TR^2}$, versus normalized displacement, $\frac{\delta}{u_o}$, for different values of (b) $C_\theta$ and (c) $M$. Simulation results of normalized angular velocity, $\frac{\dot{\phi}}{\dot{\phi}_{max}}$, versus time for different values of (d) $C_\theta$ and (e) $M$. (f) Design maps of the energy dissipation efficiency, $\eta$, for different combinations of $M$ and $C_\theta$. The red circles represent the data points for which the results are shown in the subsequent plots.}
\label{fig:Variability}
\end{figure}

To understand the effect of the value of $C_\theta$ and $M$ on the amount of energy dissipated, we append the computational model with a rotational damping component attached at the base of the Kresling unit (Fig. \ref{fig:Variability}a), and study the effect of varying the mass, $M$, and damping coefficient, $C_\theta$, on the energy dissipation efficiency, $\eta$. For a Kresling spring design with $u_o/R = 1.5$ and $\phi_o = 60^{\circ}$, we can see from the simulation results in Fig. \ref{fig:Variability}b that both the shape of and area under the $F-\delta$ curve change by changing the value of $C_\theta$. The area under the $F-\delta$ curve, which represents the total energy absorbed by the device, clearly increases with $C_\theta$. On the other hand, increasing the mass, $M$, increases the available energy to be absorbed without significantly altering the shape of the $F-\delta$ curve, as shown in Fig. \ref{fig:Variability}c.

Additionally, we tracked the angular velocity, $\dot{\phi}$, and as expected, found that decreasing $C_\theta$ increases the maximum velocity at the expense of slightly reducing the duration of impact (Fig. \ref{fig:Variability}d). On the other hand, increasing $M$ results in higher angular velocities and significantly shorter impact duration (Fig. \ref{fig:Variability}e). 

We then generated contour maps of the energy dissipation efficiency, $\eta$, covering the design space for 9 different combinations of $C_\theta$ and $M$, as shown in Fig. \ref{fig:Variability}f. It can be seen that there are no significant changes in the shape of the heat maps when varying $M$ and $C_\theta$; nevertheless, there are some alternations in the location and size of the "hot-zone"\footnote[2]{The term "hot-zone" is used to refer to the yellow region in the heat map, which corresponds to comparatively good performing designs with high efficiency.} between the different cases. For example, increasing $M$ tends to focus the hot-zone towards the top-left corner of the design map, whereas increasing $C_\theta$ generally reduces the size of the hot-zone, thereby filtering out relatively short and compliant designs. These designs are unable to effectively overcome the larger resisting torque emanating from the increased damping. As such, taller and relatively stiffer designs (top-left corner of the design map) are capable of operating at larger damping levels, in turn, absorbing large amount of the imparted (input) energy as strain energy before directing the remaining energy into rotation. Therefore, increasing the mass, $M$, increases the amount of available input energy to be dissipated through rotation, utilizing the full potential of such designs. Ultimately, the optimum performing design will, to some extent, depend on the mass and damper being used in this system.

\section{Energy Dissipation Module}
\label{sec:chamber}
\begin{figure}[t!]
\centering
\includegraphics[width=1.0\linewidth]{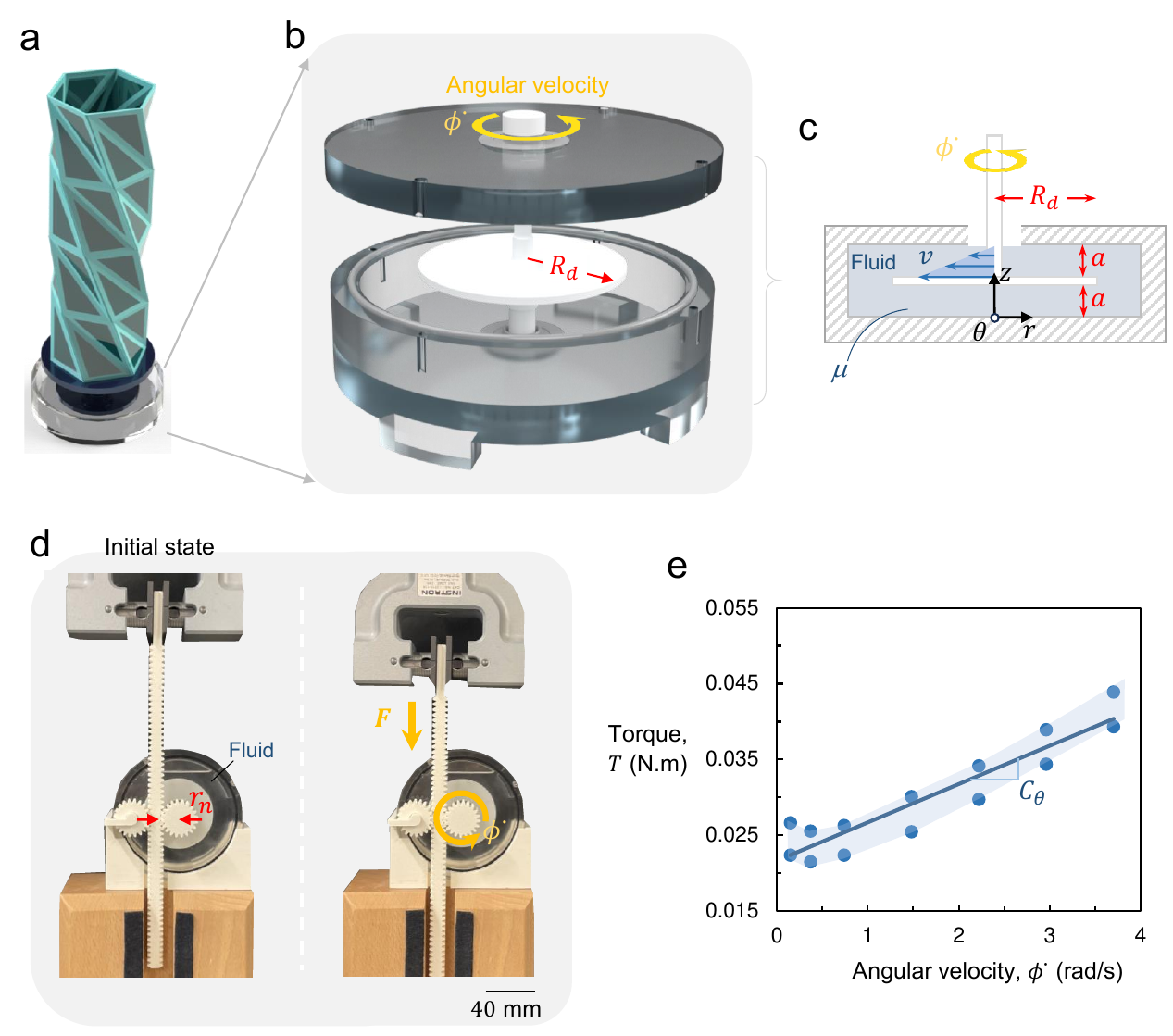}
\caption{(a) Kresling origami column with an attached rotational viscous damper at the base. (b) Viscous damping mechanism consisting of a fluid chamber and a rotating disc. (c) Schematic cross-section of the fluid chamber. (d) Experimental setup for testing the rotational damping coefficient, $C_\theta$, using uni-axial testing machine ($r_n$ is the nominal radius of the gear). (e) Experimental results of $C_\theta$ measurements. The solid line represents linear curve fitting of the experimental data points.}
\label{fig:DampCoeff}
\end{figure}

As shown in Fig. \ref{fig:DampCoeff}a, a rotational viscous damper is employed at the base of the Kresling origami column in order to dissipate the rotational energy transmitted from the load. The components of the viscous damper are illustrated in Fig. \ref{fig:DampCoeff}b, which include a rotating disc of radius, $R_d$, submerged in a fluid having a dynamic viscosity, $\mu$, and confined in a closed chamber. The torque at the base of the origami column is transmitted to the submerged disc through a connector placed at the top of the chamber and attached to a frictionless bearing.

To calculate the torque, $T$, exerted by the viscous fluid on the rotating disc, which is placed at a distance $a$ from the top and bottom walls of the chamber, as indicated in the cross-sectional schematic in Fig. \ref{fig:DampCoeff}c, we assume that the velocity profile, $v$, between the disc and the wall is linear with no-slip at the boundaries. This yields
\begin{equation}
\frac{dv}{dz}=\frac{\dot{\phi} r}{a}.
\label{eq:vel_gradient}
\end{equation}
For a Newtonian fluid, the relation between the shear stress, $\tau_{r\theta}$, acting on the plate and the velocity gradient is given by
\begin{equation}
\begin{split}
\tau_{r\theta} = \mu\frac{dv}{dz}= \mu\frac{\dot{\phi} r}{a}.\\
\end{split}
\label{eq:fluid_shear}
\end{equation}
Since the shear stress is acting on the top and bottom surfaces of the disc, we can calculate the torque  as follows:
\begin{equation}
\begin{split}
T = \int 2\tau_{r\theta} r dA =  \frac{4\pi \mu }{a} \dot{\phi} \int_{0}^{R_d} r^3 dr= \frac{\pi \mu R_d^4 }{a}\dot{\phi}.
\end{split}
\label{eq:fluid_torque}
\end{equation}
The interaction between the disc and the fluid can be modeled as a linear viscous damper, such that $T=C_\theta \dot{\phi}$, where $C_\theta$ is the rotational damping coefficient. This yields
\begin{equation}
C_\theta = \frac{\pi \mu R_d^4 }{a}.
\label{eq:fluid_damping}
\end{equation}
Since the instantaneous power generated by any arbitrary torque, $T$, is simply $T \dot{\phi}$, the energy dissipated in the viscous fluid through the rotational motion can be written as:
\begin{equation}
E_v = \int_{0}^{t_f} C_\theta \dot{\phi}^2(t) dt ,
\label{eq:fluid_energy}
\end{equation}
where $\dot{\phi}(t)$ is the angular velocity of the base throughout the compressive motion of the Kresling origami column, $t$ is the time parameter, and $t_f$ is the final time at full compression.

We can see in equation (\ref{eq:fluid_damping}) that the rotational damping coefficient can be increased by increasing the viscosity of the fluid, the radius of the disc, or reducing the gap between the disc and the chamber's walls.  In the experiments, we use Silicone Oil as the viscous fluid with a viscosity of $10$~Pa.s, and a disc of radius $R_d = 30$~mm at a distance $a = 8$~mm from the walls. Using equation (\ref{eq:fluid_damping}), the rotational damping coefficient is calculated as $C_\theta = 0.0032$~Nms/rad. To validate this result, $C_\theta$ was also obtained experimentally using a uniaxial testing machine while employing rack and pinion gears, as shown in Fig. \ref{fig:DampCoeff}d. The rack and pinion serve as a mechanism to convert the axial force and displacement into torque and rotation, respectively. Here, we measure the generated torque while varying the prescribed angular velocity, from which we obtain the rotational damping coefficient, $C_\theta$, as the slope of the $T-\dot{\phi}$ curve in Fig. \ref{eq:fluid_damping}e. Using this experimental scheme, we found $C_\theta = 0.0051$~Nms/rad, which is larger than the value calculated theoretically. The discrepancy between the theoretically calculated and experimentally measured values of $C_\theta$ could be attributed to the losses due the interaction between the fluid and the rod connecting the disc to the origami column. Such effects were ignored in the theoretical calculation. Additionally, the gap, $a$, may not be small enough to assume a linear profile for the shear stress across the gap distance.

\section{Experimental Evaluation and Performance}
\label{sec:results}
\begin{figure}[t!]
\centering
\includegraphics[width=1.0\linewidth]{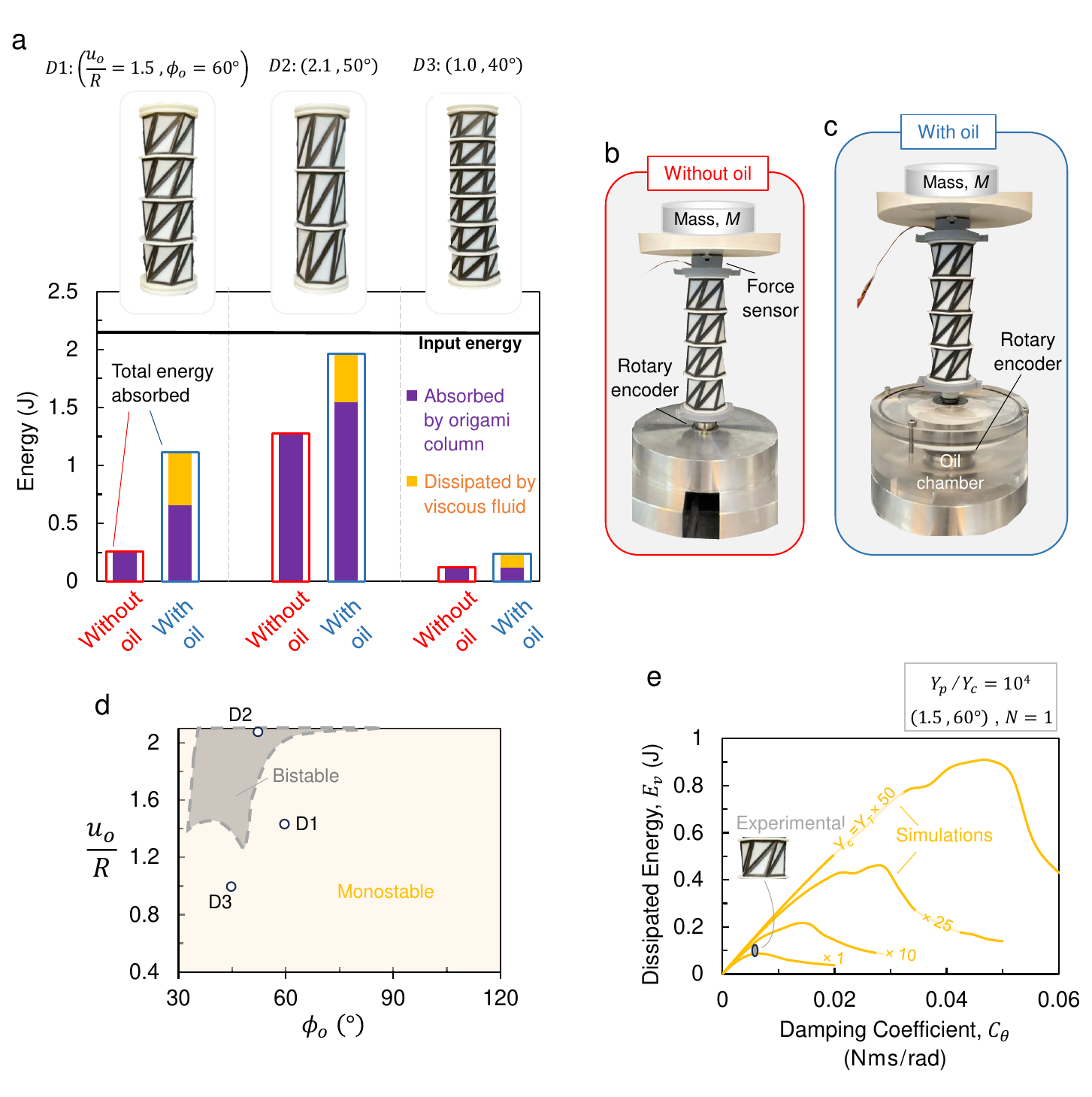}
\caption{(a) Energy absorbed by the three different designs, $D1$, $D2$, and $D3$. (b) Experimental setup without oil. (c) Experimental setup with an oil chamber at the base. (d) Bifurcation map showing regions of monostable and bistable designs. (e) Simulation results of the variability study on the materials Young's moduli. $Y_p$ and $Y_c$ are the Young's modulus of the panels and creases, respectively. The grey marker indicates experimental results for a Kresling unit made from Tango and Vero materials. }
\label{fig:ExpEval}
\end{figure}

In order to evaluate the energy dissipation performance of the proposed system, we fabricated and experimentally tested several Kresling origami columns of three different designs (refer to \ref{app:fab} for details on the fabrication process). In these experiments, we employed a mass, $M$, of $2.45$~kg and, as aforementioned in Section \ref{sec:chamber}, the rotational damping coefficient, $C_\theta$, of the energy dissipation module was measured as $0.005$~Nms/rad. The first design, $D1:$ ($u_o/R = 1.5,~\phi_o = 60^{\circ}$), and second design, $D2:$ ($u_o/R = 2.1,~\phi_o = 50^{\circ}$), were chosen from the hot-zone of the design map, while the third design, $D3:$ ($u_o/R = 1.0,~\phi_o = 40^{\circ}$), was chosen from the region of under-performing Kresling units, for comparison. The number of units in each column was selected such that the total height is similar across all columns, as shown in Fig. \ref{fig:ExpEval}a. This permits validating the effect of the efficiency parameter, $\eta$, on  performance. 

Two tests were performed for each design; one with and one without the oil chamber, from which we measure the force, $F$, displacement, $\delta$, angular velocity, $\dot{\phi}$, over time (see Supplementary Movie 1 for a high-speed recording of the origami column experiment). Then, the total energy absorbed by the system was calculated from the area under the $F-\delta$ curve ($\int{F}d\delta$). Part of this energy is stored as elastic strain energy in the materials of the origami column, and another part is dissipated by the viscoelasticity of these materials. In the presence of the oil chamber, there is also a segment of the absorbed energy being dissipated by rotations in the viscous fluid, and was calculated as $\int{C_\theta \dot{\phi}^2}dt$.

The results of the three sets of experiments are depicted in Fig. \ref{fig:ExpEval}a. Since, for all tests, the employed mass, $M$, and the total height of each column are the same, the input energy is the same and is calculated as $E_{in} = 2.2 \pm 0.05$~J. For column $D1$, the energy absorbed by the origami column in the absence of the viscous fluid was measured at around $0.26\pm 0.05$ J, and at around $0.66 \pm 0.05$ J with the oil chamber. This increase can be attributed to the increase in the restoring torque required to overcome the viscosity of the fluid and to the change in the velocity profile of the column when the viscous fluid is added. The total energy absorbed by the column and the viscous fluid was measured at $1.11 \pm 0.05$ J, which is an increase by 327\% compared to the energy absorbed in the absence of oil chamber. Portion of the absorbed energy is dissipated by viscoelasticity, whereas the amount that is dissipated by the viscous fluid was measured as $0.45 \pm 0.05$~J, leading to a dissipation efficiency of $\eta \approx 20\%$ in the fluid.

The second design, $D2$, is capable of absorbing much higher amounts of energy when compared to $D1$ mostly through the origami column, but less so through the viscosity of the fluid. For instance, the energy absorbed by the column in the absence of the viscous fluid is approximately $1.28 \pm 0.05$~J and increases to approximately $1.55 \pm 0.05$~J when the fluid is added. The energy dissipated by the viscous fluid is slightly lower than $D1$ and is measured at about $0.41 \pm 0.05$~J for a dissipation efficiency of $\eta \approx 19\%$, in addition to the energy dissipated by viscoelasticity. Since the value of $\eta$ is close between $D1$ and $D2$, the energy dissipated by viscous fluid per unit height is nearly the same for both designs.
Finally, as expected, the third design, $D3$, chosen from the low-efficiency region of the design map, demonstrated poor performance with and without the viscous fluid attaining an energy dissipation efficiency of $\eta \approx 5\%$.

The input energy that was presented earlier is a theoretical value that is calculated based on the design height ($Nu_o$) of the origami column; i.e. $E_{in} = MgNu_o$. However, the actual energy that the system is subjected to should be based on the maximum displacement ($\delta_{max}$) that the origami column have compressed until reaching densification, such that $E_{in} = Mg\delta_{max}$. For $D1$ and $D2$ columns, $\delta_{max}$ was measured as 70.1 mm and 79.6 mm, and accordingly, the actual input energy is calculated as $1.68\pm 0.05$~J and $1.91\pm 0.05$~J, respectively. This leads to an actual dissipation efficiency in viscous fluid of 27\% and 21.5\% for $D1$ and $D2$ columns, respectively. Additionally, the total energy absorbed by the $D2$ column in the presence of the viscous fluid was $1.9\pm  0.05$~J, indicating that it was able to fully absorb the imparted energy.

When comparing $D1$ and $D2$ columns, we can see that the total energy absorbed by $D2$ is higher; however, it is mainly dominated by the viscoelastic and strain energy in the constituent materials, whereas the effect of dissipating energy through rotation in oil is slightly more prevalent in $D1$. Therefore, the best design choice for a certain application depends on the type of material used to fabricate the Kresling spring column and its energy dissipation capability. For example, if the main mode of dissipation is through base rotation rather than the material, then $D1$ will be more suitable. On the other hand, if the material used has high energy dissipation capacity, then $D2$ is more suitable. It is worth mentioning that, column $D1$ behaves as a monostable spring, while $D2$ has a bistable behavior, as shown on the bifurcation map in Fig \ref{fig:ExpEval}d. A column made from bistable Kresling units does not possess the ability to recover its shape after being subjected to compressive impact loading, unlike a monostable column which can fully recover its shape upon the removal of the load (see Supplementary Movie 2 for a demonstration of a monostable and a bistable design). 
Hence, a $D2$ column is suited for single-use applications, where the bellow can be used as a sacrificial element that is disposable after impact, while $D1$ can be employed in applications requiring a recoverable device that is reusable after several successive impacts. Nevertheless, both columns, $D1$ and $D2$, were able to compress by more than 75\% of their full height before reaching densification, clearly highlighting the compressibility of the proposed energy dissipation device.

All of the previous results were obtained for Kresling springs made from Tango and Vero materials using the fabrication process described in \ref{app:fab}. Nevertheless, a Kresling spring can be fabricated from different, more resilient, materials to satisfy the requirements of more heavy-duty applications. A stiffer spring should be able to induce rotations at higher values of damping coefficient, $C_\theta$; and therefore, the choice of constituent materials has a significant effect on the energy dissipation performance. In order to study this effect, a series of simulations were carried out, in which the Young's moduli of the panel and crease materials were varied. The ratio of Young's modulus between panel and crease materials was kept constant, so as to preserve the qualitative restoring force behavior of the Kresling spring \cite{DALAQ2022110541}. Additionally, the dropped mass was set to a high value of $M = 100$~kg to be able to compress soft and stiff springs alike. Simulation results are plotted in Fig. \ref{fig:ExpEval}e, which indicate that energy dissipation increases with increasing $C_\theta$, and peaks at a point beyond which the dissipated energy starts decreasing again due to the inability of the spring to generate enough torque to overcome the damping torque. As Young's moduli are increased, the spring becomes stiffer and able to overcome larger resisting torques caused by larger values of $C_\theta$, pushing the rotational energy dissipation capacity to higher values.

\section{Conclusions}
\label{sec:conc}

We presented, optimized, and tested a novel energy absorber that employs a Kresling origami column capable of seamlessly transforming incident loads, impacts, and sudden arbitrary shocks to localized rotational energy, which, in turn, is dissipated via an energy dissipation module. The dissipation module adopted here consisted of a thin disk immersed in a cylindrical chamber filled with a viscous fluid (silicone oil). Along with the viscosity of the fluid, the dimensions of the disk and chamber determine the effective damping coefficient, $C_\theta$, of the energy dissipation mechanism.  

Apart from the chamber and fluid's geometric and material parameters, the angular velocity profile of the disc plays an important role in maximizing the amount of dissipated energy (recall $E_v = \int{C_\theta\dot{\phi^2}dt}$). This, in turn, depends on the geometric and material parameters of the origami column, which can be optimized to maximize energy dissipation for given loading conditions.  For a load in the form of a falling mass, $M$, it was observed that the design parameters ($u_o/R, \phi_o$) of the origami column leading to highest dissipation efficiency depend on the values of $M$ and $C_\theta$. A parametric study of the theoretical dissipation efficiency of the module as function of ($u_o/R, \phi_0$) reveals that there is a region (hot-zone) in the parameters' space where the efficiency is maximized. This region is localized near high values of $u_o/R$ and small values of $\phi_o$. For a given mass, the size of this region decreases as the $C_\theta$, is increased. It was shown that an optimal module designed to absorb all the imparted energy can dissipate 40\% of the absorbed energy in the viscous fluid, while the rest is either dissipated by the viscoelasticity of the origami column or stored in it as potential energy.

Finally, the outcome of these experiments do not represent the absolute maximum potential of the presented system. Materials with higher elastic moduli can be employed in the fabrication method of the origami column to produce stiffer responses capable of interacting with extreme values of $C_\theta$ and $M$. The main parameter to maintain is a sufficient contrast between the Young's moduli of panels and creases, i.e., $Y_p/Y_c > 10^4$.

\section*{Acknowledgments}
\label{sec:ack}
This research was carried out using the Core Technology Platform resources at New York University Abu Dhabi. In particular, we would like to acknowledge the support of the Advanced Manufacturing Core. Also, this material was partially supported by Tamkeen under NYUAD RRC Grant No. CG011.

\setcounter{figure}{0}
\section*{Supplementary Materials}
\appendix

\section{Details of Computational Model}
\label{app:comp}
\begin{figure}[t!]
\centering
\includegraphics[width=1\linewidth]{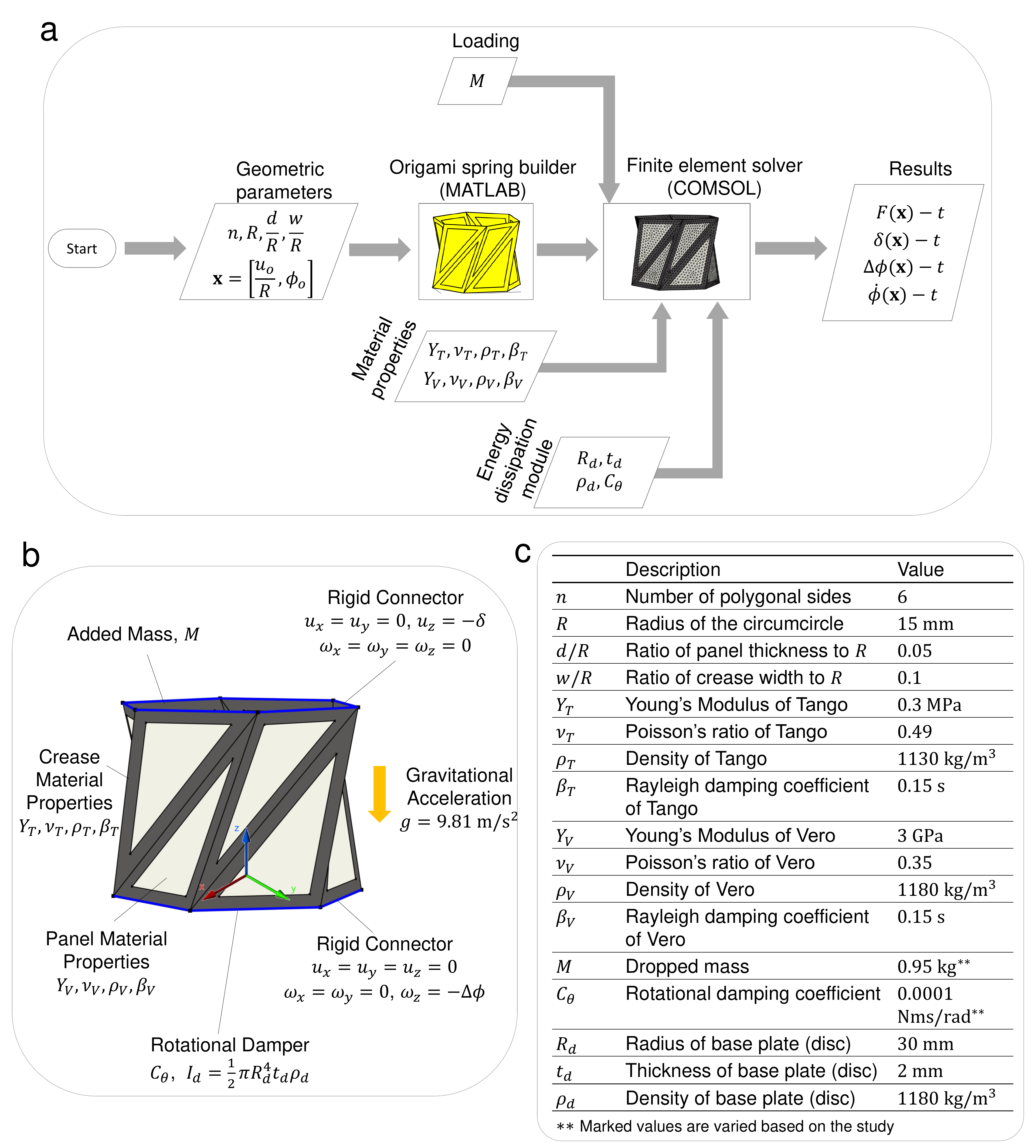}
\caption{(a) Flow chart of the automated process for computational modeling. (b) Boundary conditions applied to the FEA COMSOL model. $I_d$ is the mass moment of inertial of the base disc. (c) Geometric and material properties used in the simulations and in the actual fabricated device. }
\label{fig:FEAchart}
\end{figure}

Computational modeling was carried out by finite element analysis (FEA) using COMSOL Multiphysics\textsuperscript{\textregistered}. The model consists of a dynamic (time-domain) simulation involving a mass, $M$, dropped on top of the Kresling spring, and obtaining the energy absorption and dissipation parameters resulting from the applied load. In order to study the spring's performance over the entire design space, a large number of simulations need to be performed; hence, we developed an automated process by integrating COMSOL Multiphysics\textsuperscript{\textregistered} with MATLAB\textsuperscript{\textregistered} and utilizing parallel computing. This process starts with feeding the geometric design parameters ($n,~R,~d/R,~w/R,~u_o/R,~\text{and}~\phi_o$) to a MATLAB code that generates the Kresling structure, sends it to a prebuilt COMSOL model, and runs the simulation. The design space is created by sweeping $u_o/R$ from 0.5 to 2.5 with a step of 0.1, and $\phi_o$ from 35$^{\circ}$ to 105$^{\circ}$ with a step of 2.5$^{\circ}$, resulting in a total of 609 simulations. The prebuilt model contains all the boundary conditions, material properties, and energy dissipation module parameters necessary to run the simulations. This entire process is controlled through MATLAB, which extracts the results after the simulations are done, and performs the required post-processing to generate the contour maps. Figure \ref{fig:FEAchart}a shows a flowchart of this process.

In the prebuilt COMSOL model, the solver is based on the shell mechanics interface, which assumes that in-plane strains vary linearly along the thickness, and stresses are negligible in the transverse direction. A rigid connector boundary condition is applied to the top and bottom ends of the Kresling structure. This boundary condition emulates the behavior of having rigid bodies attached to the structure. The top end is fixed in rotation but free to translate along the $z-$axis, that is $u_x=u_y=\omega_x=\omega_y=\omega_z=0$ and $u_z=-\delta$ \footnote[3]{$~u_x$ , $u_y$, and $u_z$ are the $x$, $y$, and $z$ components of the translational displacement field in Cartesian coordinates, respectively. $\omega_x$, $\omega_y$, $\omega_z$ are the $x$, $y$, and $z$ components of the rotational (angular) displacement field, respectively.}, while the bottom end is fixed in translation but allowed to rotate freely about the $z-$axis; i.e. $u_x=u_y=u_z=\omega_x=\omega_y=0$ and $\omega_z=\Delta\phi$, as shown in Fig. \ref{fig:FEAchart}b. An added mass boundary condition is applied at the top end, and a gravity force condition is included to drive the mass, $M$, at $g=9.81~\text{m/s}^2$ gravitational acceleration. Moreover, a rotational damper is attached to the base (bottom end) incorporating inertial and viscous effects of the energy dissipation module. The elastic and inertial material properties are defined for Tango and Vero, which are the crease and panel materials, respectively. In addition, material damping is included through the stiffness-proportional damping coefficient, $\beta$, of the Rayleigh damping model \cite{priestley2005viscous}, which moderately takes into account the viscoelasticity in the material, and aids in solver's convergence. The values of all simulation parameters are tabulated and provided in Fig. \ref{fig:FEAchart}c.

Since simulating contact mechanics consumes heavy computational resources, we did not include it in the computational model to be able to run the large number of simulations efficiently. Nevertheless, we added two criteria to stop the simulation at the time instance when the creases and panels are interacting, which mimics the point where the Kresling structure reaches densification, and minimizes numerical artifacts. The first criterion, $C_{r1}$, is based on the locking of the Kresling spring that occurs due to the intersection of all creases at a single point at large rotation angle, $\phi$, exceeding $(\frac{n-1}{n})\times180^{\circ}$ \cite{DALAQ2022110541}. The first stopping criterion can be written as:
\begin{align*}
C_{r1}: \phi > \left(\frac{n-1}{n}\right)\times180^{\circ}-TOL_1 \\
\text{For } n=6 \rightarrow C_{r1}: \phi > 150^{\circ}-TOL_1
\end{align*}
The second criterion, $C_{r2}$, stops the simulation when the spring reaches the maximum achievable compression height; i.e. when the displacement, $\delta$, exceeds the initial height, $u_o$, minus the thickness of interacting panels during a certain rotation angle range, as follows:
\begin{equation*}
\text{For } n=6 \rightarrow C_{r2}:
    \begin{cases}
      \delta > u_o-4d\times TOL_2 & \text{for $30^{\circ}\le\phi<60^{\circ}$}\\
      \delta > u_o-6d\times TOL_2 & \text{for $60^{\circ}\le\phi<150^{\circ}$}
    \end{cases} 
\end{equation*}
where the factors ``$4$" and ``$6$" are the number of interacting panels. $TOL_1$ and $TOL_2$ are tolerances to accommodate for panels/creases thickness and other unconsidered effects such as panels/creases bending and twisting, and their values were set as $20^{\circ}$ and $1.2$, respectively.

\section{Fabrication}
\label{app:fab}
\begin{figure}[t!]
\centering
\includegraphics[width=0.8\linewidth]{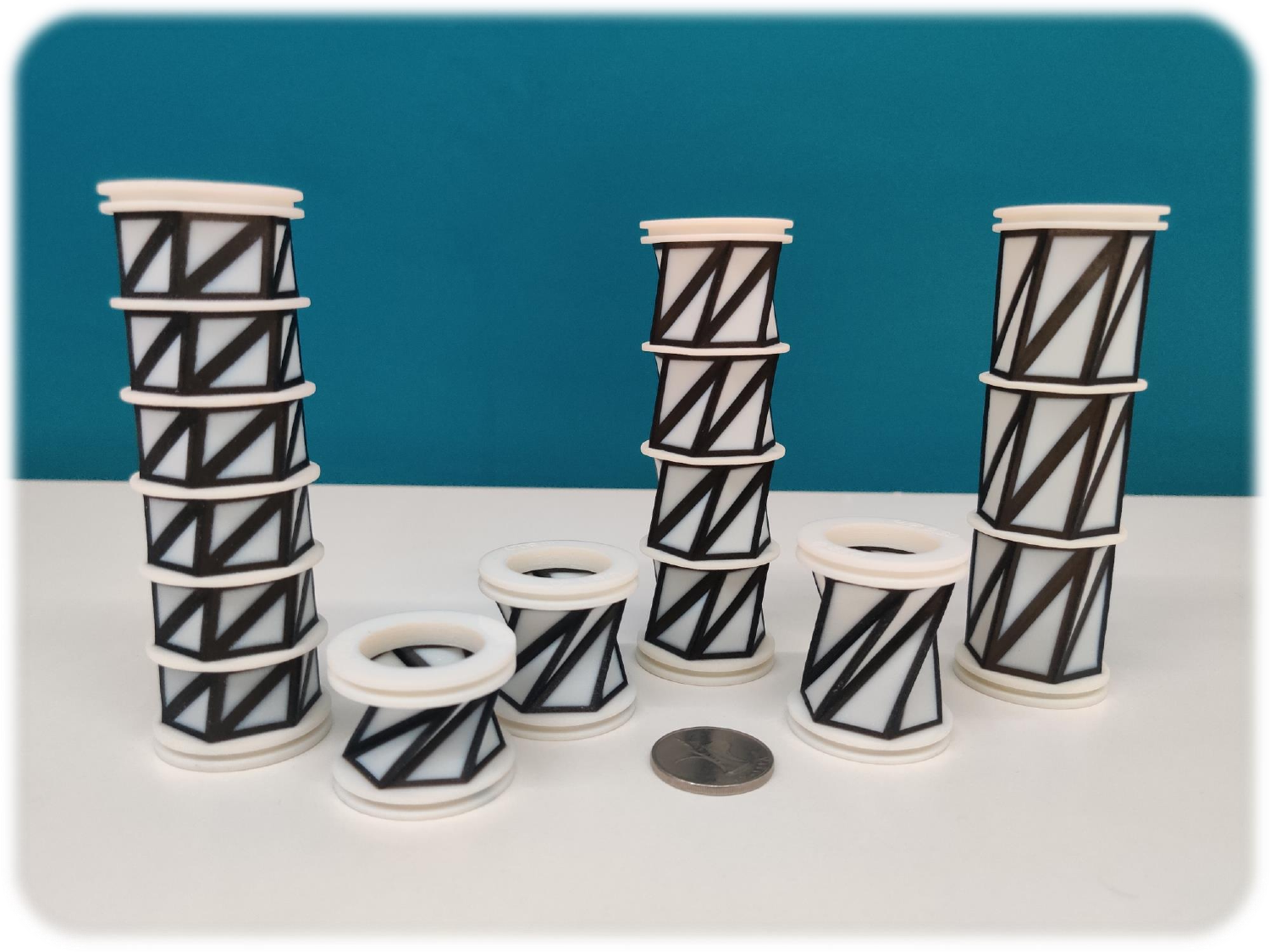}
\caption{3D printed samples of Kresling spring units and columns. }
\label{fig:samples}
\end{figure}

The Kresling spring units and columns are fabricated by utilizing the polyjet 3D printing technology of the Stratasys J750 3D printer. The structures comprise a composite of flexible rubber-like material (TangoBlackPlus \cite{stratasys_tango_2018}), and a hard stiff material (Vero \cite{stratasys_vero_2018}). The flexible material is used to create the creases, which represent the outer frame of the triangular elements in order to permit stretching and bending during deformation, whereas the stiff material is used for the panels to allow them to fold while retaining the shape and functionality of the Kresling origami pattern during deformation. After 3D printing, each sample is submerged in a chemical solution of 2\% caustic soda (sodium hydroxide NaOH) and 1\% sodium metasilicate (Na$_2$SiO$_3$) for a period ranging from 18 to 72 hours, depending on the size of the sample, to ensure complete removal of the SUP706 support material \cite{stratasys_find_2022}. The samples are then rinsed and cured under UV-light for 300-500 minutes. Figure \ref{fig:samples} shows some 3D printed Kresling spring units and columns.

The main components of the energy dissipation module are an oil chamber and a submerged rotating disc. The oil chamber is constructed by machining of acrylic blocks, and the rotating disc is 3D printed from Vero material. All the geometric parameters of the fabricated Kresling spring samples and the rotating disc are listed in a table in Fig. \ref{fig:FEAchart}c.

\section{Experimental Setup}
\label{app:exp}
The experimental testing setup is built in-house and consists of a free-falling impactor head guided via two frictionless bearings along two steel poles. The impactor head is placed on top of the KOS sample while its bottom end is attached to an optical rotary encoder with a resolution of 1000 pulses per revolution (Fig. \ref{fig:validation}a). The mass of the impactor head can be adjusted by adding different weights via a bolt-nut mechanism. The applied mass in the tests of single Kresling spring units is $0.95$~kg, whereas the mass is set to $2.45$~kg in the stacked columns tests. A laser sensor (Micro-Epsilon optoNCDT 2300) with a precision of $\pm 20~\mu m$ is used to track the position of the impactor head along the vertical $z$-axis during compression, while a force sensor is used to measure the reaction force after releasing the impactor. In the stacked columns experiments, an oil chamber portraying the energy dissipation module is employed at the base. The obtained data (reaction force, impactor position, rotation angle and time) are collected via an acquisition system with a sampling rate of 20 kHz. The data is used to extract the force-time ($F-t$), deflection-time $(\delta-t$), angular rotation-time ($\Delta\phi-t$) curves, from which we generate the force-deflection ($F-\delta$) and angular velocity-time ($\dot{\phi}-t$) curves and calculate the absorbed and dissipated energies.

\section*{Data availability}
The raw/processed data required to reproduce these findings cannot be shared at this time as the data also forms part of an ongoing study.

\bibliographystyle{ieeetr} 
\bibliography{Main}





\end{document}